\documentclass[12pt]{iopart}

\usepackage{iopams}
\usepackage{graphicx}
\expandafter\let\csname equation*\endcsname\relax
\expandafter\let\csname endequation*\endcsname\relax
\usepackage{amsmath}
\usepackage{bm}
\usepackage{hyperref}
\usepackage{graphics}
\usepackage{slashed}
\usepackage{amssymb}
\usepackage{url}
\usepackage[dvipsnames,svgnames,table]{xcolor}
\usepackage[T1]{fontenc}

\makeatletter
\newcommand{\tpitchfork}{%
  \vbox{
    \baselineskip\z@skip
    \lineskip-.52ex
    \lineskiplimit\maxdimen
    \m@th
    \ialign{##\crcr\hidewidth\smash{$-$}\hidewidth\crcr$\pitchfork$\crcr}
  }%
}
\makeatother

\usepackage[normalem]{ulem}
\usepackage{color}

\begin{document}

\title{Metric-Induced Principal Symbols in Nonlinear Electrodynamics}
\author{\'Erico Goulart${}^{1}$ and Eduardo Bittencourt${}^{2}$\footnote{Corresponding author}}
\address{${}^{1}$Federal University of S\~ao Jo\~ao d'El Rei, C.A.P. Rod.: MG 443, KM 7, CEP-36420-000, Ouro Branco, MG, Brazil}
\address{${}^{2}$Federal University of Itajub\'a, Av. BPS 1303, Pinheirinho, Itajub\'a-MG, 37500-903, Brazil}
\ead{egoulart@ufsj.edu.br, bittencourt@unifei.edu.br}

\date{\today}

\begin{abstract}
We present a geometrical formulation of nonlinear electrodynamics by expressing its principal symbol as an optical metric-induced object. Under the assumption of no birefringence, we show that the evolution of linear perturbations can be recast as a covariant divergence on a curved, field-dependent background, enabling the application of quantum field theory techniques as in Maxwell's theory in curved backgrounds. This geometric reformulation opens a new route for analog models potentially implementable in laboratory via tailored nonlinear metamaterials.
\end{abstract}

\vspace{2pc}

\section{Introduction}
Analog models of gravity based on fluid dynamics have gained considerable prominence in recent years, both theoretically and experimentally \cite{novello2002artificial,barcelo2005,Jacquet2020,Field2021,Braunstein2023}. These systems allow for the construction of analogies that mimic quantum field theory in curved spacetimes, enabling the simulation of phenomena such as Hawking radiation \cite{Unruh1981}, ``trans-Planckian modes'' \cite{Jacobson1991} and their prospective detection in analogue platforms \cite{Fischer2017,Fischer2023,Fischer2025}, as well as Bekenstein–Hawking entropy \cite{VisserPRL1998}, among others \cite{Visser1998,barcelo2005}. Notable realizations include experiments involving vortices \cite{Fischer2002,Oliveira2015,Torres2017,Patrick2018,Torres2020,Svancara2024,smaniotto2025}, superfluid helium \cite{Finne2006,Svancara2024}, Bose–Einstein condensates \cite{Anderson2013,Steinhauer2016}, and related media.

While there exists a vast literature, from different research groups, on analog models derived from nonlinear electrodynamics (NLED) \cite{AYONBEATO1999,Novello2000,DELORENCI2000,Hehl2003,Novello2004,DUNNE_2005,breton2007nonlinear,Goulart2009,Abalos:2015gha,Sorokin_2022,alam2022review,Goulart2024,Goulart2025,Goulart2025IJGMMD}, these models have not achieved the same level of maturity as those from fluid dynamics. One fundamental reason lies in the richer structure of the electromagnetic field, which possesses more degrees of freedom than scalar excitations in fluids. Moreover, the quasi-linear nature of the field equations leads to perturbations governed by a more intricate algebraic structure, confining analogies with curved spacetime to the kinematic level.

In this letter, we show that this limitation can be overcome. We identify sufficient conditions under which the evolution equation for linear perturbations in NLED can be recast as a covariant divergence on an effective curved spacetime determined by a fixed background field. This reformulation opens a new pathway for exploring quantum aspects of NLED using techniques akin to those employed in linear electrodynamics on curved backgrounds. Although the resulting expression resembles that of Maxwell’s theory in curved spacetime, the reformulation is far from trivial. Establishing the existence of such decomposition for the principal symbol---especially in nonlinear and birefringent contexts---remains an open problem \cite{Gilkey2001,Schuller2006}. Our result shows that, under the physically motivated condition of non-birefringence, the principal symbol admits a non-obvious but exact decomposition in terms of a single metric.

Finally, this theoretical advance is supported by recent experimental developments. Nonlinear metamaterials have been proposed as platforms to simulate effective geometries with tunable refractive indices~\cite{Ciattoni2010,Alu2011,Smolyaninov2010metric}, allowing for the emulation of analog gravity effects in the electromagnetic sector and potentially bridging quantum field theory in curved spacetimes with engineered nonlinear media.

\section{Principal Symbol in Nonlinear Electrodynamics}\label{princ_NLED}
We consider a four-dimensional spacetime $(M, g_{ab})$ with signature $(+,-,-,-)$ and focus on gauge-invariant, nonlinear theories of electrodynamics governed by the action
\begin{equation}
S = \int L(\psi, \phi) \sqrt{-g} \, \mathrm{d}^4 x,
\end{equation}
where the scalar invariants are defined by
\begin{equation}
\psi \equiv \frac{1}{2} F^{ab} F_{ab}, \qquad \phi \equiv \frac{1}{2} \star F^{ab} F_{ab}.
\end{equation}
Here, $F_{ab}$ is the electromagnetic field strength and $\star F^{ab}$ its Hodge dual. Varying the action yields the system of quasi-linear field equations
\begin{equation}
\label{eq:field_eq}
\nabla_b \left( L_\psi F^{ab} + L_\phi \star F^{ab} \right) = 0, \qquad \nabla_b \star F^{ab} = 0,
\end{equation}
where $L_\psi$ and $L_\phi$ denote partial derivatives of the Lagrangian with respect to $\psi$ and $\phi$, respectively.

Assuming $L_\psi \neq 0$, a necessary (but not sufficient) condition for hyperbolicity\footnote{Hyperbolicity is at the roots of physics, since it amounts to the predictable power of the theory, asserting that solutions exist, are unique and depend continuously on the initial data. We refer the reader to \cite{Abalos:2015gha}, where necessary and sufficient conditions for symmetric-hyperbolicity are discussed in details.}, the first equation above can be recast as
\begin{equation}
\label{eq:NLED-dyn}
X^{abcd} \, \nabla_b F_{cd} = 0,
\end{equation}
where

\begin{equation}
\label{princ_symbol}
X^{abcd} \equiv g^{a[c}g^{bd]} + \xi_1 F^{ab} F^{cd} + \xi_2 \left( F^{ab} \star F^{cd} + \star F^{ab} F^{cd} \right) + \xi_3 \star F^{ab} \star F^{cd},
\end{equation}
with coefficients defined as
\begin{equation}
\label{eq:def_xi}
\xi_1 \equiv \frac{2 L_{\psi\psi}}{L_\psi}, \qquad
\xi_2 \equiv \frac{2 L_{\psi\phi}}{L_\psi}, \qquad
\xi_3 \equiv \frac{2 L_{\phi\phi}}{L_\psi}.
\end{equation}
The tensor $X^{abcd}$ is called the \emph{Principal Symbol} of the system of partial differential equations and controls the main qualitative features of the theory, such as the characteristic surfaces, domains of influence, and evolution of small electromagnetic disturbances \cite{John2014,Obukhov2002}. Clearly, it is antisymmetric in the pairs $ab$ and $cd$ and symmetric under their mutual exchange. 

Accounting for linear perturbations $\delta F_{ab}$ about a given smooth background solution $F_{ab}$, equation (\ref{eq:NLED-dyn}) gives:
\begin{equation}
\label{eq:NLED-dyn_pert}
\nabla_b\left(X^{ab}{}_{cd} \delta F^{cd}\right) +Y^{a}{}_{cd}\,\delta F^{cd} = 0,
\end{equation}
with
$$Y^{a}{}_{cd}=\left( \frac{\delta X^{ab}{}_{pq}}{\delta F^{cd}} - \frac{\delta X^{ab}{}_{cd}}{\delta F^{pq}} \right)\nabla_{b}F^{pq}.$$
In this form, the dispersion relation governing the perturbations is not manifest, nor it is clear which background geometry should be employed for quantization. Though rather intricate, the second term in equation (\ref{eq:NLED-dyn_pert}), being algebraic in $\delta F_{ab}$, does not influence the characteristic structure and therefore plays no essential role in the determination of wave propagation.

\section{Effective Metrics in NLED}
It is well known that, in the geometric limit of equation (\ref{eq:NLED-dyn_pert}), the wave fronts are governed by the characteristic polynomial
\begin{equation}\label{charpol}
Y^{ac} (^{\star} X^{\star}_{abcd}) Y^{bd} = 0,
\end{equation}
where $^{\star} X_{abcd}^{\star}$ is the double dual, $Y^{ac} \equiv X^{abcd} k_b k_d$ is the characteristic matrix, and $k_a$ is the characteristic covector. A general framework for extracting the covariant dispersion relation from the principal symbol was developed in~\cite{Itin2009}, where the characteristic surfaces are derived directly from the constitutive tensor without assuming a background metric. In general, this quartic expression factorizes into two quadratics, indicating birefringence: distinct polarizations propagate along different light cones.

However, for a special class of Lagrangians, birefringence is absent. This occurs when the fully-nonlinear relations hold \cite{Obukhov2002}:
\begin{eqnarray}
(\xi_1 \xi_3 - \xi_2^2)\psi &=& \xi_1 - \xi_3, \label{biref1}\\
(\xi_1 \xi_3 - \xi_2^2)\phi &=& 2\xi_2. \label{biref2}
\end{eqnarray}
The Born–Infeld model is the most extensively studied example of this class, although other non-birefringent models also exist. Indeed, a recent work has classified all solutions of these conditions for general NLED, identifying not only Born–Infeld, but also Plebanski, and additional exceptional models that satisfy the same criteria~\cite{Russo2023}. In all such cases, the quartic surface degenerates into a single effective light cone
\begin{equation}
(h^{-1})^{ab}k_a k_b = 0,
\end{equation}
where the \textit{inverse effective metric} can be written as
\begin{equation}
\label{eff_met_Ob_Rub}
(h^{-1})^{ab} \equiv \mathcal{X} g^{ab} - \mathcal{Y} t^{ab},
\end{equation}
with $t^{ab} \equiv F^a_{\phantom{a}k} F^{kb}$, $
\mathcal{X}\mathcal{Y} \equiv \xi_3$, and $\mathcal{Y} \equiv (\xi_1 \xi_3 - \xi_2^2)^{1/2}$. Here, the conformal factor was purposely chosen to match the so-called Boillat metric, first discussed in \cite{Boillat1970}. As described in \cite{ Gibbons2001}, this choice has several peculiar features and makes the determinant of the effective metric coincide with that of the background metric. 

In what follows, we shall prove a remarkable result connecting the principal symbol of every non-birefringent theory with the effective metric determined by the background field configuration. In other words, we shall demonstrate that if equations (\ref{biref1}) and (\ref{biref2}) hold, then
\begin{equation}
X^{abcd} = (h^{-1})^{a[c} (h^{-1})^{bd]}.
\end{equation}
It is worth mentioning that such a splitting is a particular realization of the so-called Gilkey’s decomposition theorem, first discussed in \cite{Gilkey2001}. Roughly speaking, the theorem guarantees that any algebraic curvature map can be written as a linear combination of smaller blocks of algebraic curvature maps that are induced from a finite collection of non-degenerate tensors. However, since no constructive algorithm for the decomposition is currently known (see \cite{Schuller2006} for details), it is rather surprising that the single block presented above is constructed precisely with the Boillat metric. In this vein, our result is rather nontrivial due to algebraic subtleties and, from a physical point of view, implies that this class of nonlinear theories somehow echo the behavior of Maxwell fields in curved spacetimes.

\section{Decomposing the principal symbol}

In this section, we shall decompose all relevant tensors with respect to a field of observers. As usual, we consider a smooth future-oriented unit timelike vector field, say $v^{a}(x)$, and use it to split the corresponding tensors into their respective spatial and temporal quantities. In particular, the Faraday tensor splits as
\begin{equation}\label{Faraday_split}
F^{ab}=E^{[a}v^{b]}+\varepsilon^{ab}_{\phantom a\phantom a cd}B^{c}v^{d}
\end{equation}
with $E^{a}=F^{a}_{\phantom a b}v^{b}$ and $B^{a}=-\star F^{a}_{\phantom a b}v^{b}$
denoting the electric and magnetic fields as measured by the observers. Similarly, an arbitrary double-form splits as \cite{Goulart2025IJGMMD}
\begin{equation}\label{Beldec}
X_{abcd}=\{g_{abpq}(g_{cdrs}\mathbb{A}^{pr}+\varepsilon_{cdrs}\mathbb{B}^{pr})+\varepsilon_{abpq}(g_{cdrs}\mathbb{C}^{pr}+\varepsilon_{cdrs}\mathbb{D}^{pr})\}v^{q}v^{s},
\end{equation}
with\footnote{We stick to the convention where the position of the star denotes the pair of antisymmetric indices which is to be Hodge dualized.}
\begin{equation}\label{Beldec1}
\mathbb{A}_{ac}=X_{abcd}v^{b}v^{d},\quad \mathbb{B}_{ac} =-X^{\star}_{abcd} v^{b}v^{d},\quad\mathbb{C}_{ac}= -^{\star}X_{abcd} v^{b}v^{d},\quad \mathbb{D}_{ac} =\ ^{\star}X^{\star}_{abcd} v^{b}v^{d}.
\end{equation}
Clearly, $E^{a}$, $B^{a}$ and the tetrad $\{\mathbb{A}_{ac},\mathbb{B}_{ac},\mathbb{C}_{ac},\mathbb{D}_{ac}\}$ are all spacelike and 
orthogonal to the observers by construction.

According to equations (\ref{Faraday_split}) and (\ref{Beldec}), the observers split the principal symbol provided by Eq. (\ref{princ_symbol}) as
\begin{eqnarray}
\mathbb{A}_{ac}&=&+h_{ac}+\xi_{1}E_{a}E_{c}-\xi_{2}E_{(a}B_{c)}+\xi_{3}B_{a}B_{c},\\
\mathbb{B}_{ac}&=&+\xi_{1}E_{a}B_{c}+\xi_{2}(E_{a}E_{c}-B_{a}B_{c})-\xi_{3}B_{a}E_{c},\\
\mathbb{C}_{ac}&=&+\xi_{1}B_{a}E_{c}+\xi_{2}(E_{a}E_{c}-B_{a}B_{c})-\xi_{3}E_{a}B_{c},\\
\mathbb{D}_{ac}&=&-h_{ac}+\xi_{1}B_{a}B_{c}+\xi_{2}E_{(a}B_{c)}+\xi_{3}E_{a}E_{c},
\end{eqnarray}
where $h_{ab}=g_{ab}-v_{a}v_{b}$ is the usual projector. We notice that $\mathbb{A}_{ac}$ and $\mathbb{D}_{ac}$ are symmetric tensors, whereas the quantities $\mathbb{B}_{ac}$ and $\mathbb{C}_{ac}$ do not have a well defined symmetry.

Now, consider an auxiliary tensor, defined by the anti-symmetrized product
\begin{equation}
M^{abcd}\equiv M^{a[c}M^{bd]}
\end{equation}
where $M^{ac}\equiv \epsilon g^{ac}+\delta t^{ac}$ with free functions $\epsilon$ and $\delta$ to be determined. In this case, the observer splits this object into a tetrad of tensors $\{\mathbb{P}_{ac},\mathbb{Q}_{ac},\mathbb{R}_{ac},\mathbb{S}_{ac}\}$, in the same way as the principal symbol was decomposed, namely 
\begin{equation}
\mathbb{P}_{ac}\equiv M_{abcd}v^{b}v^{d},\quad \mathbb{Q}_{ac}\equiv-M^{\star}_{abcd} v^{b}v^{d},\quad\mathbb{R}_{ac}\equiv-^{\star}M_{abcd} v^{b}v^{d},\quad \mathbb{S}_{ac}\equiv\ ^{\star}M^{\star}_{abcd} v^{b}v^{d}.
\end{equation}
A tedious algebraic calculation yields
\begin{eqnarray}
\fl\qquad\mathbb{P}_{ac}=\left(\epsilon^{2}-\epsilon\delta\psi-\frac{\delta^{2}\phi^{2}}{4} \right)h_{ac}+\left(\delta^{2}\psi-\epsilon\delta\right)E_{a}E_{c}-\frac{\delta^{2}\phi}{2}E_{(a}B_{c)}-\epsilon\delta B_{a}B_{c},\\
\fl\qquad\mathbb{Q}_{ac}=(\delta^{2}\psi-\epsilon\delta)E_{a}B_{c}+\delta^{2}\frac{\phi}{2}(E_{a}E_{c}-B_{a}B_{c})+\epsilon\delta B_{a}E_{c},\\
\fl\qquad\mathbb{R}_{ac}=(\delta^{2}\psi-\epsilon\delta)B_{a}E_{c}+\delta^{2}\frac{\phi}{2}(E_{a}E_{c}-B_{a}B_{c})+\epsilon\delta E_{a}B_{c},\\
\fl\qquad\mathbb{S}_{ac}=-\left(\epsilon^{2}-\epsilon\delta\psi-\frac{\delta^{2}\phi^{2}}{4} \right)h_{ac}-\epsilon\delta E_{a}E_{c}+\frac{\delta^{2}\phi}{2}E_{(a}B_{c)}+(\delta^{2}\psi-\epsilon\delta) B_{a}B_{c}.
\end{eqnarray}

Let us suppose that the principal symbol of a NLED may be written in the form of
\begin{equation}
X^{abcd}=M^{abcd},
\end{equation}
with the tensors defined as above. At first sight, our assumption seems impossible because we have only two free functions ($\epsilon$ and $\delta$) and a large number of equations, given by
\begin{equation}
\mathbb{P}_{ac}=\mathbb{A}_{ac},\quad \mathbb{Q}_{ac}=\mathbb{B}_{ac},\quad \mathbb{R}_{ac}=\mathbb{C}_{ac},\quad \mathbb{S}_{ac}=\mathbb{D}_{ac}.
\end{equation}
However, by comparing the appropriate tensor splittings, one realizes that several relations are redundant, and we end up with the set of equations
\begin{eqnarray}
\fl\qquad \epsilon^{2}-\epsilon\delta\psi-\frac{\delta^{2}\phi^{2}}{4}=1,\qquad
\xi_{1}=\delta^{2}\psi-\epsilon\delta, \qquad \xi_{2}=\frac{\delta^{2}\phi}{2}, \quad \mbox{and} \quad
\xi_{3}=-\epsilon\delta.
\end{eqnarray}
Simple algebraic manipulations then reveal that the solution of the system is given by
\begin{equation}\label{Boillatmet}
\epsilon=\xi_{3}/(\xi_{1}\xi_{3}-\xi_{2}^{2})^{1/2},\quad\quad\quad \delta=-(\xi_{1}\xi_{3}-\xi_{2}^{2})^{1/2},
\end{equation}
provided that
\begin{eqnarray}\label{Boillatrel}
(\xi_{1}\xi_{3}-\xi_{2}^{2})\psi=\xi_{1}-\xi_{3},\quad\quad\quad(\xi_{1}\xi_{3}-\xi_{2}^{2})\phi=2\xi_{2}.
\end{eqnarray}
Equations (\ref{Boillatmet}) reveal that the auxiliary tensor $M^{ab}$ must coincide with the Boillat metric (\ref{eff_met_Ob_Rub}), whereas equations (\ref{Boillatrel}) imply differential constraints that must be satisfied by Lagrangian, surprisingly coinciding with the no-birefringence conditions (\ref{biref1})-(\ref{biref2}). \hfill$\Box$

\section{Perturbation dynamics}
The results presented so far have a striking consequence for analog models of gravity constructed with NLED. Indeed, since the principal symbol of every non-birefringent theory decomposes as a Kulkarni-Nomizu product of the effective metric with itself, electromagnetic perturbations $\delta F^{ab}$ about a background solution $F^{ab}$ will behave as if they were embedded in an effective curved spacetime. 

The proof proceeds very much in the same way as it does in the context of fluids. Start by constructing a covariant derivative $\widetilde{\nabla}_a$ compatible with the covariant effective metric, say $h_{ab}$. Now, recalling the fact that the Boillat form (\ref{eff_met_Ob_Rub}) defines the same element of volume as the background metric, i.e. $\det(h_{ab})=\det(g_{ab})$,
simple calculations using equation (\ref{eq:NLED-dyn_pert}) reveal that
\begin{equation}
(h^{-1})^{ac}(h^{-1})^{bd}\, \widetilde{\nabla}_b\, \delta F_{cd} + \frac{1}{2}Y^{a}{}_{cd} \, \delta F^{cd} = 0.
\end{equation}
In other words, the evolution equation for perturbations may be written in terms of a covariant divergence constructed with the effective metric evaluated on top of the given background solution. Needless to say, this result permits the application of standard techniques akin to quantum field theory in curved spacetime in the context of NLED, opening a new pathway of investigation within the analog gravity program.

In quantum field theory, particularly in curved or effective spacetimes, the principal symbol plays a central role in determining the \textit{hyperbolicity} and \textit{causal structure} of the equations of motion. These properties are not merely mathematical: they directly constrain the form of Green’s functions, the well-posedness of the Cauchy problem, and ultimately the construction of the Hilbert space of states. In a more general context, the role of the principal symbol in determining a consistent quantization procedure has been investigated in Ref.\ \cite{Rivera2011}, where general linear electrodynamics is quantized through the identification of a well-defined hyperbolic structure from the constitutive tensor. That construction relies on the ability to derive a covariant Hamiltonian formulation and consistent vacuum states, hinging critically on the causal structure induced by the principal symbol. Moreover, since the characteristic surfaces (or light cones) are encoded in the principal symbol, the geometric reinterpretation of the latter makes manifest the conditions under which a consistent quantization is viable.

\section{Connecting with experiments}

The metric-induced decomposition of the principal symbol establishes a direct and
model-independent link between NLED and the electromagnetic response of structured media. For a generic gauge-invariant NLED described by an action, the field equations (\ref{eq:field_eq}) can always be written in the
form
\begin{equation}
\nabla_b H^{ab} = 0,
\qquad
\nabla_b \star F^{ab} = 0,
\label{eq:medium_form}
\end{equation}
where the excitation tensor
\begin{equation}
H^{ab} \equiv L_\psi F^{ab} + L_\phi \star F^{ab}
\label{eq:excitation_tensor}
\end{equation}
plays the role of a covariant constitutive relation. In this form, NLED is naturally
interpreted as electrodynamics in a nonlinear medium whose response depends on the
background electromagnetic field configuration.

This observation provides a systematic bridge between NLED and analog gravity
setups based on nonlinear optical media. In contrast to most analog models, which
are restricted to the geometric optics regime, the results presented in this work apply
to the \emph{full wave operator} governing linear perturbations. Indeed, when the
non-birefringence conditions are satisfied, the principal symbol factorizes as a
Kulkarni--Nomizu product of an effective inverse metric,
\begin{equation}
X^{abcd} = (h^{-1})^{a[c}(h^{-1})^{bd]},
\label{eq:metric_factorization}
\end{equation}
and the linearized equations of motion can be recast in the covariant divergence form
derived in Sec.~5. As a consequence, the effective metric $h_{ab}$ controls not only
the characteristic surfaces, but the entire dynamical evolution of electromagnetic
perturbations around a given background solution.

Nonlinear metamaterials have emerged as promising candidates to emulate effective geometries in electrodynamics. Several experimental platforms have already demonstrated the ability to engineer field-dependent refractive indices and strong nonlinearities. Plasmonic nanorod arrays, for instance, exhibit nonlinear refractive index enhancements exceeding two orders of magnitude~\cite{Capretti2015}, enabling the study of wave propagation in regimes where the electromagnetic field cannot be approximated by its geometric limit. Similarly, epsilon-near-zero (ENZ) media are known to exhibit dominant higher-order nonlinearities, with third-, fifth-, and even seventh-order terms playing a significant role~\cite{Reshef2017}. In these regimes, the dynamics of perturbations become sensitive to the full structure of the wave operator, precisely the quantity controlled by the principal symbol. These systems have been used to realize spatially modulated nonlinear propagation, dynamic light trapping, and analog gravitational lensing~\cite{Grinblat2021,Gennaro2022}, suggesting that the implementation of effective curved metrics is within reach.

Our findings provide a theoretical framework for systematically identifying which nonlinear media can support such analog simulations beyond the kinematic level. The known mapping between NLED Lagrangians and material media via field-dependent permittivity and permeability tensors, developed in Ref.\ \cite{Delorenci2014}, becomes particularly powerful in the non-birefringent case, allowing the translation of a chosen NLED model (such as Born-Infeld) into a concrete constitutive relation for engineered media. To this end, we propose the following roadmap: one begins by selecting a physically motivated NLED model with a non-birefringent structure. Using the established constitutive correspondence, one translates the derivatives of the Lagrangian into effective dielectric and magnetic responses. These can then be implemented using nonlinear photonic media such as ENZ platforms, metasurfaces, or multi-layered structures designed to respond nonlinearly to background fields. Crucially, wave propagation experiments in such media---under fixed background field configurations---can probe features like modified causal cones, horizon-like boundaries, and signature transitions, directly emulating the full dynamics predicted by NLED.

\section{Discussion}
We have shown that, under the assumption of vanishing birefringence, NLED admits a dynamical reformulation in which the principal symbol factorizes in terms of an effective metric. This identification promotes the optical geometry, usually confined to geometric optics, to the full dynamical level of the theory, bringing the NLED closer to the status of analog gravity models in fluid systems.

The viability of quantization in such effective geometries has been previously demonstrated in general linear theories~\cite{Rivera2011}, where causal structure and hyperbolicity were used to derive consistent Hamiltonian dynamics and vacuum states. Our results suggest that similar constructions are now possible in nonlinear settings under suitable conditions, potentially extending the reach of quantum field techniques into broader electromagnetic media. The algebraic nature of this decomposition also raises interesting mathematical questions regarding the classification of $X^{abcd}$ and its reducibility, which remain largely unexplored.

On the experimental side, our decomposition suggests the possibility of emulating nonlinear field dynamics through suitably engineered media. The effective metrics derived from NLED share structural similarities with those obtained from dielectric and metamaterial setups. This correspondence enables a direct mapping between theoretical NLED models and physical realizations in nonlinear optics.

One of the most intriguing open questions is whether similar geometric decompositions can be extended to birefringent models, where the principal symbol splits into multiple metrics. Ultimately, our results suggest that NLED may serve as a bridge between fundamental theory and laboratory-based simulations of emergent spacetimes.

\appendix

\section{A special example: Born--Infeld electrodynamics}
\label{sec:born-infeld}

In order to make the discussion more concrete, by exhibiting an explicit nonlinear model for which all steps of the construction can be carried out analytically, in this section we therefore work out the Born--Infeld Lagrangian as a fully explicit example. In particular, we compute the coefficients $\xi_i$ entering the principal symbol (\ref{princ_symbol}), verify the non-birefringence conditions (\ref{biref1})--(\ref{biref2}), and display the corresponding effective metric (\ref{eff_met_Ob_Rub}) and covariant constitutive relation, thereby providing a direct bridge to experimental/analog implementations discussed in the previous section.

We consider the Born--Infeld model written as a function of the Lorentz invariants, and a convenient parametrization as
\begin{equation}
\label{eq:BI_L}
L_{\rm BI}(\psi,\phi)=b^{2}\left(1-\sqrt{\Delta}\right),\qquad
\Delta \equiv 1+\frac{\psi}{b^{2}}-\frac{\phi^{2}}{4b^{4}},
\end{equation}
where $b$ sets the nonlinearity scale and $\Delta>0$ ensures that the Lagrangian is real. Expanding this Langragian at weak fields ($|\psi|,|\phi|\ll b^{2}$) gives that Born--Infeld reduces to Maxwell theory in the appropriate limit.

From equation (\ref{eq:BI_L}) one finds the first derivatives
\begin{equation}
\label{eq:BI_first}
L_{\psi}=-\frac{1}{2\sqrt{\Delta}},\qquad 
L_{\phi}=\frac{\phi}{4b^{2}\sqrt{\Delta}},
\end{equation}
and the second derivatives
\begin{equation}
\label{eq:BI_second}
\fl L_{\psi\psi}=\frac{1}{4b^{2}}\Delta^{-3/2},\qquad
L_{\psi\phi}=-\frac{\phi}{8b^{4}}\Delta^{-3/2},\qquad
L_{\phi\phi}=\frac{1}{4b^{2}}\Big(1+\frac{\psi}{b^{2}}\Big)\Delta^{-3/2}.
\end{equation}
These closed-form expressions already display the two key features relevant for propagation: (i) $L_{\psi}\neq 0$ as long as $\Delta>0$, and (ii) the nonlinearities enter through the same scalar $\Delta$ that will also control the effective cone.

Using the definitions given by equation (\ref{eq:def_xi}), the Born--Infeld coefficients take the explicit form
\begin{equation}
\label{eq:BI_xi}
\xi_{1}=-\frac{1}{b^{2}\Delta},\qquad
\xi_{2}=\frac{\phi}{2b^{4}\Delta},\qquad
\xi_{3}=-\frac{1}{b^{2}\Delta}\Big(1+\frac{\psi}{b^{2}}\Big).
\end{equation}
A straightforward computation then yields
\begin{equation}
\label{eq:BI_disc}
\xi_{1}\xi_{3}-\xi_{2}^{2}=\frac{1}{b^{4}\Delta}.
\end{equation}
With (\ref{eq:BI_xi})--(\ref{eq:BI_disc}) one can verify explicitly that the non-birefringence conditions (\ref{biref1})--(\ref{biref2}) are satisfied identically:
\begin{equation}
(\xi_1\xi_3-\xi_2^{2})\,\phi=\frac{2\phi}{b^{4}\Delta}=2\xi_2,
\end{equation}
and
\begin{equation}
(\xi_{1}\xi_{3}-\xi_{2}^{2})\,\psi=\frac{\psi}{b^{4}\Delta}=\xi_{1}-\xi_{3}.
\end{equation}
Therefore, Born--Infeld belongs to the special class of theories for which the characteristic polynomial degenerates into a \emph{single} light cone, i.e.\ the two polarizations propagate on the same effective geometry determined by (\ref{eff_met_Ob_Rub}).

When equations (\ref{biref1})--(\ref{biref2}) hold, the inverse effective metric can be written as (\ref{eff_met_Ob_Rub}), whose coefficients are
\begin{equation}
\label{eq:BI_Y_and X}
\fl\qquad\mathcal{Y}=\sqrt{\xi_{1}\xi_{3}-\xi_{2}^{2}}
=\frac{1}{b^{2}\sqrt{\Delta}},\qquad \mathcal{X}=\frac{\xi_3}{\mathcal{Y}}
=-\frac{b^{2}+\psi}{b^{2}\sqrt{\Delta}}.
\end{equation}
Hence, the Born--Infeld effective metric is obtained in closed form as
\begin{equation}
\label{eq:BI_effmet}
(h^{-1})^{ab}
=-\frac{1}{2b^2\sqrt{\Delta}}\Big(2\,t^{ab}+(b^{2}+\psi)\,g^{ab}\Big),
\end{equation}
where an overall conformal factor is physically irrelevant for the characteristic cone $(h^{-1})^{ab}k_{a}k_{b}=0$, but its choice is crucial to obtain the Boillat metric. Finally, one can verify that $\det(h_{ab})=\det(g_{ab})$ directly for Born--Infeld.


\section*{Acknowledgements}
We are thankful to J. P. Pereira and G.B. Santos for the valuable discussions and comments, helping to improve the manuscript. EB is partially supported CNPq (grant N.\ 305217/2022-4)

\section*{References}

\bibliography{ref.bib}

\end{document}